\documentstyle[12pt]{article}
\begin{document}
\newcommand{\bea}{\begin{eqnarray}}
\newcommand{\eea}{\end{eqnarray}}
\renewcommand{\thefootnote}{\fnsymbol{footnote}}
\newcommand{\be}{\begin{equation}}      \newcommand{\ee}{\end{equation}}
\newcommand{\st}{\scriptsize}
\newcommand{\fs}{\footnotesize}

\vspace{1.3in}
\begin{center}
\baselineskip 0.3in
{\bf  Klein Paradox for bound states - A puzzling phenomenon}

\vspace {0.1in}
{Nagalakshmi A. Rao}

\vspace{-0.2in}
{\it Department of Physics, Government Science College,}

\vspace{-0.2in}
{\it Bangalore-560001, Karnataka, India}

\vspace{-0.2in}
{\it drnarao@gmail.com}

\vspace{0.1in}
{B.A. Kagali}

\vspace{-0.2in}
{\it Department of Physics, Bangalore University,}

\vspace{-0.2in}
{\it Bangalore-560056, Karnataka, India}

\vspace{-0.2in}
{\it bakagali@hotmail.com}
\end{center}
\hspace{2.3in}
{\bf Abstract}

\indent
While Klein paradox is often encountered in the context of scattering of relativistic particles
at a potential barrier, we presently discuss a puzzling situation that arises with the
Klein-Gordon equation for bound states. With the usual minimal coupling procedure of introducing
the interaction potential, a paradoxical situation results when the "hill" becomes a "well",
simulating a bound state like situation. This phenomenal phenomenon for bound states is
contrary to the conventional wisdom of quantum mechanics and is
analogous to the well-known Klein paradox, a generic property of relativistic wave equations.

\vspace{0.0in}
\noindent

\vspace{0.3in}
\hspace{-0.27in}
PACS NOs.: 03.65.Ge, 03.65.Pm
\newpage
\indent

\section{\bf \sf \sc Introduction}
\setcounter{equation}{0}
\hspace{0.1in}\, In non-relativistic quantum mechanics, the scattering of an electron by a potential barrier is known to be one of the
simplest solvable problems. However, a similar problem with a potential step or a barrier in relativistic quantum mechanics,
often results in paradoxical situations, called the Klein Paradox.

\indent
In the original work of Klein{$^1$}[1929], electrons incident on a large potential step was addressed. The problem was treated with the Dirac
equation and it was found that for large potentials, $V(x)>E+mc^{2},$ the reflection coefficient, $R_{S}$, exceeds unity
while the transmission coefficient, $T_{S}$, becomes negative. This suggested that more particles are refelected by the step
than are incident on it. Such a puzzling situation contradicting non-relativistic expectation was termed Klein Paradox.

\indent
Several authors{$^{2-6}$} have discussed Klein Paradox under various circumstances. Considering
a potential step with sharp boundaries,
Bjorken{$^7$} has illustrated that a weak potential having decaying exponential solutions inside the potential region
leads to undamped oscillatory solutions for potentials exceeding $(E+mc^{2})$, consistent with the original version of Klein
Paradox. However, Greiner{$^8$}, on the basis of the group velocity treatement has illustrated an unexpected largness of the
transmission coefficient. While Brojken's explanation of Klein Paradox is essentially based on pair production, Greiner's
representation is that of single-particle interpretation.

\indent
Similar results exist for Klein Gordon particles as well. Guang Jiong et.al{$^9$}. have shown that the Klein-Gordon equation with a step potential in minimal coupling exhibits the Klein Paradox
at the one-particle level. Rubin Landau{$^{10}$} has given a reasonable explanation of the Klein Paradox based on
particle-antiparticle pair production.

\vspace{-0.1in}
\indent
In the following section, we discuss the appearance of a paradoxical result in the context of
bound states.

{\section{\bf \sf \sc The Klein-Gordon Equation With a Potential}}
\setcounter{equation}{0}
\indent
The time-independent one-dimensional Klein-Gordon equation for a general potential introduced as a vector field, may well be
written as
\bea
\left\{ {d^{2}\over
dx^{2}}+{\left(E-V\left(x\right)\right)^{2}-m^{2}{c}^{4}\over {c}^{2}{\hbar
}^{2}}\right\} \psi (x)=0.
\eea
This equation may be cast into the form of Schrodinger equation as
\bea
{d^{2}\psi \over d{x}^{2}}+{2m\over {\hbar
}^{2}}\left(E_{\!eff}-{V\!}_{eff}\right)\psi \left(x\right)=0,
\eea
with
\bea
E_{\!eff}={E^{2}-{m}^{2}{c}^{4}\over 2m{c}^{2}}
\eea
and
\bea
V_{\!\!eff}={2E\ V(x)-V^{2}\left(x\right)\over 2mc^{2}}.
\eea
\indent
The concept of effective energy and effective potential used to simulate the properties of
relativistic wave equations leads to paradoxical results.

\vspace{-0.1in}
\indent
We first review the finite square well problem and then address the case of potential hill.

\noindent
{\large \bf A.  Potential Well}

\indent
For an attractive square well potential defined by
\bea
V\left(x\right)=\left\{ \matrix{-V_{0}\ \ \ {\rm for}\ \ \left|x\right|\leq
a\cr \cr \ 0\ \ \ \ {\rm for}\ \ \ \left|x\right|>a,} \right.
\eea
the effective potential takes the form
\bea
V_{\!\!eff}\left(x\right)={-2E\ V_{0}\ -\ V_{0}^{2}\over
2mc^{2}}\ \ {\rm for}\ \left|x\right|\leq a
\eea
and vanishes elsewhere. Considering positive energy states, the effective potential looks like another square well potential
and therefore, so long as $E_{\!eff}<0$ or $E<mc^{2}$, bound states are possible. This result is quite reasonable.

\noindent
{\large \bf B.  Potential Hill}

\indent
We now consider a potential barrier defined by
\bea
V\left(x\right)=\left\{ \matrix{+V_{\!0}\ \ \ {\rm for}\ \ \ \left|x\right|\leq
a\ \cr \cr 0\ \ \ \ {\rm for}\ \ \ \left|x\right|>a,} \right.
\eea
which is not the usual Klein step, but has better-defined boundaries.
The effective potential for a `hill' takes the form
\bea
V_{\!\!eff}\left(x\right)={2E\ V_{0}\ -\ V_{0}^{2}\over 2mc^{2}},
\eea
\indent
the effective energy remains the same. It is trivial to check that
\vspace{0.1in}
\bea
\left(E_{\!eff}-V_{\!\!eff}\right)=\left\{
\matrix{{E^{2}-m^{2}{c}^{4}+V_{0}^{2}-2EV_{0}\over 2mc^{2}} & \ \ \ {\rm for} \ \ \  &
\left|x\right|\leq a\cr\cr {E^{2}-{m}^{2}{c}^{4}\over {2mc}^{2}} & \ \ \ {\rm for} \ \ \
& \left|x\right|>a} \right.
\eea
\indent
Interestingly, $V_{\!\!eff}$ can be positive, zero or even negative for a range of values of the
barrier height $V_{0}$.

\indent
For $V_{0}<2mc^{2}$, $V_{\!\!eff}$ remains positive, and the problem is analogous to a typical scattering
problem.

\indent
For $V_{0}=2mc^{2}$, the effective potential vanishes and the barrier becomes supercritical.

\indent
However a puzzling situation arises for a barrier height exceeding $2mc^{2}$. As is seen from
Eqn.(8), the effective potential becomes negative and thus a `hill' is transformed into a
`well'. For $(V_{0}>2mc^{2})$, particles, instead of being scattered by the potential "hill" are trapped inside the simulated "well".
This means that bound states are possible for strong barriers. Such a paradoxical situation
may be called the Klein Paradox for bound states.

\indent
  Moreimportantly, it may be inferred that
the potential hill need not be only of the square type for such
anomalous bound states. The actual number of the bound states and the energies, however, will
depend on the shape and size of the hill. These trivial results may be worked
out from the standard procedures of quantum mechanics.

\section{\bf \sf \sc Results and Discussion}
\setcounter{equation}{0}
\indent
While the original version of Klein Paradox concerns scattering situation of Dirac particles at a potential step, we have
shown that an analogous paradoxical situation arises even for bound states. With the
usual minimal coupling procedure of introducing interaction, the Klein-Gordon equation leads to bound states for finitely
extended potential hills, contrary to the conventional wisdom of quantum mechanics. So long as the effective vector potential
remains positive,$(V<2mc^{2})$ for repulsive potentials, the situation is similar to a typical scattering problem.
As the potential becomes stronger and stronger, exceeding the limiting value $E+mc^{2}$
the `hill' becomes a `well' simulating a bound state-like situation.Surprisingly, a typical scattering problem is transformed into
a bound state problem. Thus {\it Klein Paradox is retrieved for
bound states in the case of a strong repulsive, finite ranged barrier}.

\indent
Interestingly, a paradoxical situation like this does not arise for pure scalar repulsive potentials. In such a case, a
barrier is transformed into another barrier, no matter how strong or weak the potential is.
The paradoxical result for bound state illustrates that
Klein-Gordon equation is reasonable even at the one particle level.

\vspace{0.3in}
\noindent
{\large \bf Acknowledgements}

One of the authors (N.A.Rao) is grateful to the Director of Collegiate Education in Karnataka, 
Prof. K.V.Kodandaramaiah  for his support. Thanks are extended to Prof. T. Gangadaraiah, 
Principal, Govt. Science College for his encouragement.
\newpage
\baselineskip 0.2in
\hspace{2.3in}
{\Large \bf References}
\begin {enumerate}

\item Klein O.Z. {\it Physics}, {\bf 53} (1929) 157-165.

\item Mark J.Thomson and Bruce H.J.McKellar, {\it The solution of the Dirac equation for high squre barrier}, Am. J. Phys. {\bf 59}(4) (1991) 340-346.

\item Francisco Dominguez-Adame, {\it A relativistic interaction without Klein Paradox}, Phys. Lett. {\bf A 162} (1992) 18-20.

\item Holstein B.R., {\it Klein Paradox}, Am.J.Phys. {\bf 69} (1999) 507-512.

\item Calogeracos A and N.Dombey, {\it Klein tunneling and Klein Paradox}, Int J. M. Phys. {\bf A 14} (4) (1999) 631-643.

\item Antonio S de Castro, {\it $(n+1)$ dimensional Dirac equation and the Klein Paradox}, Am.J.Phys. {\bf 69}(10) (2001), 1111-1112.

\item James D. Bjorken and Sidney D.Drell, {\it Relativistic Quantum Mechanics}, Mc Graw Hill Book Company, NY,(1964) pp 40-43.

\item Greiner, Walter, {\it Relativistic Quantum Mechanics-Wave Equations}, Springer-Verlag, (1995).

\item Gaung-Jiong Ni, Weimin Zhou and Jun Yan, {\it Klein Paradox and antiparticle}

\item Rubin H.Landau, {\it Quantum Mechanics II A Second Course in Quantum Theory}, A Wiley Interscience Publication, New York, 1996.

\end{enumerate}
\end{document}